\documentclass[cites,graphicx,floatfix]{article}

\usepackage{epsfig,amsmath,amssymb}
\title{Formation of ultracold dipolar molecules in the lowest vibrational levels by photoassociation}

\author{
\parbox{\textwidth}
{ J. Deiglmayr$^{a,b}$, M. Repp$^a$, A.
Grochola$^{a,}$\footnote{Also at the Institute of Experimental
Physics, Warsaw University, Poland}, K. M\"ortlbauer$^a$, C.
Gl\"uck$^a$, O. Dulieu$^b$, J. Lange$^a$, R. Wester$^a$, and M.
Weidem\"uller$^{a,}$\footnote{e-Mail: weidemueller@physi.uni-heidelberg.de} \\[3mm]
$^a$ Physikalisches Institut Albert-Ludwigs-Universit\"at Freiburg,
Germany \\[1mm]
$^b$ Laboratoire Aim\'e Cotton, University Paris-Sud XI, Orsay,
France}
}

\begin{document}
\maketitle
\renewcommand{\thefootnote}{\fnsymbol{footnote}}

\noindent We recently reported the formation of ultracold LiCs
molecules in the rovibrational ground state
X$^1\Sigma^+$,$v''$=0,$J''$=0 [J. Deiglmayr \textit{et al.}, PRL
\textbf{101}, 133004 (2008)]. Here we discuss details of the
experimental setup and present a thorough analysis of the
photoassociation step including the photoassociation line shape. We
predict the distribution of produced ground state molecules using
accurate potential energy curves combined with an ab-initio dipole
transition moment and compare this prediction with experimental
ionization spectra. Additionally we improve the value of the
dissociation energy for the X$^1\Sigma^+$ state by high resolution
spectroscopy of the vibrational ground state.

\section{Introduction}
\label{sec:intro}

Ultracold molecular gases are promising candidates for studying
diverse systems reaching from tests of the standard model to
ultracold chemistry and quantum computing. Of special interest is
the preparation of these molecules in the absolute rovibrational
ground state, as this state is stable against inelastic collisions
and allows therefore the creation of a stable, dense gas of
molecules. A number of experimental approaches are currently being
studied to prepare and manipulate ultracold molecules
\cite{doyle2004,dulieu2006}. Magneto-association using interspecies
Feshbach resonances followed by adiabatic transfer via two or more
optical transitions has been recently used to produce dense gases of
deeply bound molecules of Cs$_2$~\cite{danzl2008}, KRb~\cite{ni2008}
and Rb$_2$~\cite{lang2008}, even reaching the rovibrational ground
state. Photoassociation (PA), a long established technique, has been
successfully employed in the production of K$_2$~\cite{nikolov2000},
RbCs~\cite{sage2005}, Cs$_2$~\cite{viteau2008}, and
LiCs~\cite{deiglmayr2008b} in the lowest vibrational level of the
electronic ground state. In contrast to the homonuclear molecules,
heteronuclear alkali dimers exhibit strong dipole moments ranging
from 0.6\,Debye for LiNa and KRb to 5.5\,Debye for
LiCs~\cite{aymar2005}. The formation of these dipolar molecules in
their lowest vibrational level opens the way to the exploration of
quantum phases in dipolar gases \cite{micheli2006, pupillo2008}, the
development of quantum computation techniques \cite{rabl2006},
precision measurements of fundamental constants
\cite{zelevinsky2008}, and the investigation and control of
ultracold chemical reactions \cite{tscherbul2006}.

We recently reported the formation of ultracold LiCs molecules in
the rovibrational ground state
X$^1\Sigma^+$,$v''$=0,$J''$=0~\cite{deiglmayr2008b}. In
Sect.~\ref{sec:experiment} of this article, we discuss details of
the experimental setup with a special focus on the reduction of
interspecies losses in overlapped magneto-optical traps. After
analyzing the PA step and modeling the PA line shape in
Sect.~\ref{sec:paspec}, we focus on the distribution of populated
vibrational ground state levels in Sect.~\ref{sec:pottheory}. In the
discussion of the detection step, these predictions are compared
with experimental ionization spectra and qualitative agreement is
found (Sect.~\ref{sec:ionspec}). Finally we present high-resolution
spectroscopy of the vibrational ground state and improve the value
of the dissociation energy of the lowest X$^1\Sigma^+$ level in
Sect.~\ref{sec:depletionpec}.

\section{Experimental setup}
\label{sec:experiment}

In this section, the experimental setup for the formation and
detection of ultracold polar molecules is described. We outline the
general principles and focus on recent important additions and
improvements of the setup. Further details can be found in
references~\cite{deiglmayr2008b,kraft2006,kraft2007}.

\subsection{Trapping a large number of atoms in overlapped magneto-optical traps} \label{sec:darkspot}

\begin{figure}[htb]
  \begin{center}
   \includegraphics[width=\columnwidth,clip]{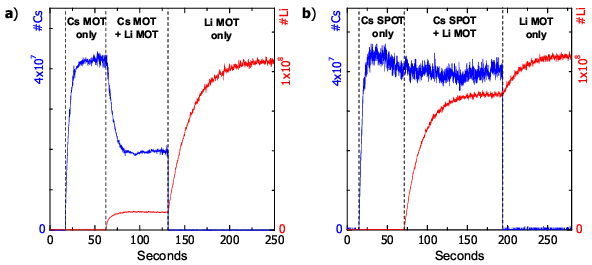}
   \caption{Loading curves for overlapped lithium and cesium traps: a) First only the cesium MOT is loaded, then additionally the lithium MOT.
   The number of trapped cesium atoms is reduced by roughly a factor of two. After blocking the cesium MOT, roughly a 9-fold increase in the number
   of trapped lithium atoms is observed, indicating very high losses due to Li-Cs collisions.
   b) The same sequence with overlapped forced dark SPOT for cesium and identical lithium MOT. Only a weak suppression of trapped atom numbers is observed
   when loading both traps simultaneously.}
   \label{fig:motvsspot}
  \end{center}
\end{figure}

% Trap setups
For the formation of dipolar ground state molecules, we cool and
trap simultaneously $^7$Li and $^{133}$Cs atoms in two overlapped
magneto-optical traps (MOTs). The atoms are evaporated in a double
species oven and are decelerated in a single Zeeman
slower\footnote{When both traps are loaded simultaneously, the
configuration of the Zeeman fields is optimized for slowing lithium
atoms. With this field configuration, the loading rate for cesium
atoms is reduced to 1x10$^7$ atoms/s which is still sufficient for
the experiments described here.}. In overlapped magneto-optical
traps for different atomic species, high loss rates due to inelastic
interspecies collisions are a well known experimental difficulty
\cite{mancini2004}. Fig.~\ref{fig:motvsspot}\,a) demonstrates the
importance of these losses for the setup used in previous
experiments~\cite{kraft2006}. The dominant loss channel has been
identified as collisions between excited Cs(6P$_{3/2}$) and ground
state Li(2S$_{1/2}$) atoms \cite{schloder1999}. It was therefore
straightforward to reduce these losses by reducing the number of
cesium atoms in the excited state. This was achieved by implementing
a dark magneto-optical trap (also called ``dark spontaneous force
optical trap'' or ``dark SPOT'' \cite{ketterle1993}) for cesium. In
the center of such a trap, the repumping light is blocked, so that
already cold atoms are pumped by the cooling light off-resonantly
into the lower hyperfine ground state which is now a dark state.
Only if they leave the central part of the trap, they are pumped
back by repumping light and take part in the cooling cycle again.
This has been widely used to increase number and density of the
trapped atoms (which we also observe), however we mainly profit from
the large fraction of atoms in the dark hyperfine ground state
(typically 97$\%$) which reduces interspecies loss-rates by the same
fraction. The repumping light in the center is blocked by imaging
the shadow of a 3.5mm large piece of plastic\footnote{It turned out
that the best extinction ratio of 1:10$^4$ at the trap position was
obtained by using the magnetic plastic from an old floppy disk.}
with an 1:1 telescope onto the trap center. As the laser beam used
for Zeeman-slowing of the atoms also contains repumping light and
passes through the center of the trap, we additionally image a dark
spot (also 3.5mm, 1:1 imaging ratio) in the Zeeman beam onto the
trap center. The Rayleigh range of the imaging focus is on the order
of only a few centimeters, so that the slowing of the atomic beam is
not significantly reduced by the shadow in the center of the Zeeman
beam. We note that the reduced absolute power in the Zeeman beam
leads to a reduced loading rate which is however more than
compensated by the reduced single species losses in a dark SPOT. In
order to reach a high dark-state fraction, we detune the repumping
laser by $\sim$5$\Gamma$ and shine an additional depumping laser on
the center of the trap (100$\mu$W, $\omega_0$=1.5mm). This setup is
called a detuned forced dark SPOT \cite{townsend1996}.
Fig.~\ref{fig:motvsspot}\,b) demonstrates the increase in the number
of simultaneously trapped lithium and cesium atoms after
implementing the detuned forced dark SPOT for cesium. With this
setup we typically trap 4x10$^7$ Cs atoms and 10$^8$ Li atoms at
densities of 3x$10^9$cm$^{-3}$ and $10^{10}$cm$^{-3}$ respectively.
Using absorption images and time-of-flight expansion we measure a Cs
temperature of 250(50)$\mu$K. The temperature of the Li atoms lies
significantly higher (hundreds of $\mu$K) due to the unresolved
hyperfine structure of the excited state and the large photon
recoil.

%Compare atom numbers between SPOT and MOT

\subsection{Molecule production and detection} \label{sec:moleculesexp}

% Ti:Sa Setup
In order to form ultracold molecules, colliding pairs of laser
cooled atoms are transferred into bound molecules by
photoassociation (PA)~\cite{jones2006}. Light from a tunable laser
is continuously shown on the overlapped atomic clouds. Two colliding
atoms can absorb a photon from this laser and form an excited bound
molecule. The spontaneous decay of these excited molecules can
either lead back to pairs of free atoms (usually with additional
kinetic energy) or into bound molecules in the lowest singlet or
triplet state. In the system described here, the latter is the
dominant process as it will be shown later
(Sect.~\ref{sec:pottheory}). The light for the PA is provided by a
commercial Ti:Sa laser (Coherent MBR-110, pumped by 9 to 12 Watts
from a Coherent Verdi V18) and is delivered to the UHV chamber by a
high power optical fibre. After the fibre, the light is collimated
to a waist of 1.0mm, matched to the size of the lithium MOT. It is
aligned to pass through the center of the overlapped cesium and
lithium MOT's by optimizing the depletion of trapped cesium atoms
with resonant light from the Ti:Sa laser. Typical laser powers after
the chamber are 400-500mW for wavelengths between 946 and 852nm. The
wavelength is measured using a home-built wavemeter, calibrated to
an atomic cesium resonance with an absolute accuracy of
0.01\,cm$^{-1}$. Additionally we monitor Ti:Sa frequency scans with
a reference cavity, which is stabilized via an offset-locked diode
laser to an atomic cesium resonance. This reference cavity is also
used for monitoring long-term drifts of the Ti:Sa laser and to lock
it to an arbitrary frequency with a remaining fluctuation of
$\lesssim$2MHz/day.

%Experimental setup for ionization
In order to detect the produced ground state molecules, they are
first ionized state-selectively by a pulsed laser. The resulting
LiCs$^+$ ions are then separated by time-of-flight mass spectrometry
from other atomic or molecular ions and are finally detected on a
microchannel plate in a single ion counting setup. For details of
the time-of-flight mass spectrometer we refer to a previous
work~\cite{kraft2006}. We only note that it allows us to clearly
separate Cs$^+$ ions from LiCs$^+$ ions which have a mass difference
of 5$\%$. For ionization, two photons of one color from a pulsed dye
laser (Radiant Dyes NarrowScan, Rhodamin B/6G pumped by 532nm,
typically 4mJ in a beam with a waist of 5mm and a pulse length of
7ns, bandwith 0.1cm$^{-1}$) are used in a resonant-enhanced
multi-photon ionization (REMPI) scheme. The pulsed laser beam passes
roughly one beam diameter below the trapped atom clouds in order to
reduce excessive ionization of cesium and lithium atoms which would
saturate the detector. Additionally the number of cesium ions is
reduced by turning off the repumping light 0.6ms before the
ionization pulse. The depumper used in the forced dark-spot setup
quickly pumps all trapped cesium atoms into the lower hyperfine
state $F$=3, so that the trapping laser is now off-resonant and the
atoms remain in the electronic ground state from where only three
photon ionization is possible. In contrast to two-photon ionization
from the excited 6P$_{3/2}$ state, this process is strongly
suppressed and therefore the possible effect on the detected
LiCs$^+$ signal is minimized. After the ionization pulse the
repumper is turned on again within 300$\mu$s and nearly all cesium
atoms are recaptured in the MOT.

\section{Photoassociation spectroscopy}
\label{sec:paspec}

\begin{figure}[tb]
  \begin{center}
   \includegraphics[width=0.6\columnwidth,clip]{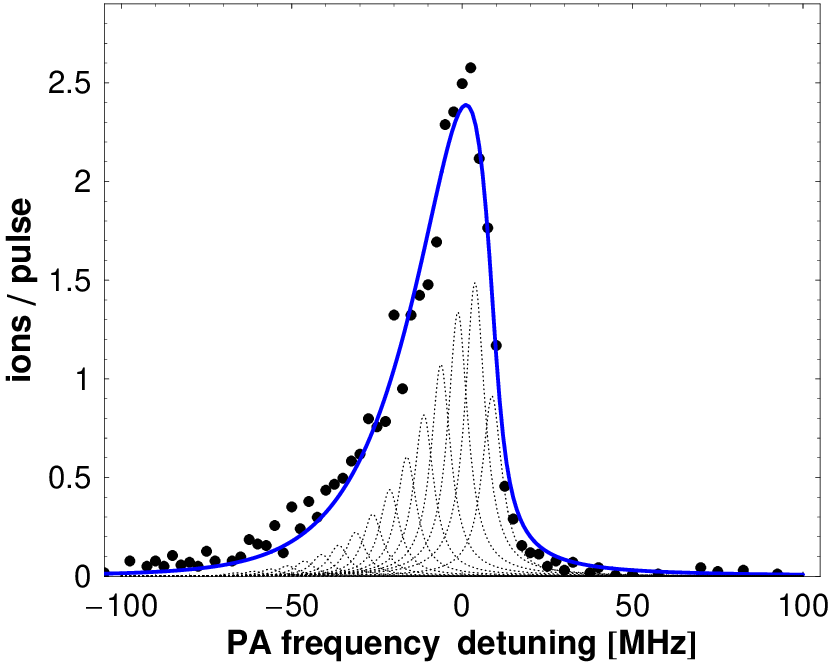}
   \caption{Temperature-broadened PA resonance (dots) together with a fit of Equation (\ref{eq:tempmodel}) (solid
   line). Best fit parameters are T$_{rel}$=580(80)$\mu$K and
   $\Gamma$=10(3)MHz. The thin dashed lines illustrate the
   contribution to the PA line shape from atom pairs with discrete collisions energies. The frequency axis shows detuning from the setpoint of
   11632.11cm$^{-1}$. The molecules are ionized by two photons at ~14693.7cm$^{-1}$.
   }
   \label{fig:tempfit}
  \end{center}
\end{figure}

%Modeling PA line shape: model
In Fig.~\ref{fig:tempfit} we show an individual resonance in a scan
of the PA laser. As the PA laser becomes resonant with a transition
from a colliding atom pair to an excited molecular level, the number
of detected LiCs$^+$ ions increases, indicating the formation of
ground state molecules. We can indeed be sure that the ion signal
arises from ground state molecules and not from excited molecules,
as the PA and ionization laser are spatially separated and the
molecules do not move significantly over the lifetime of the excited
level. We note, however, that in these first experiments the
wavelength of the ionization laser was chosen based on a rough
estimate only, relying on \textit{ab initio} potentials for the
intermediate levels and for the ion potential. The precise
ionization mechanism at wavelengths around 14700cm$^{-1}$ is still
under investigation.

The resonance shown in Fig.~\ref{fig:tempfit} does not show further
substructure, i.e. no molecular hyperfine structure, indicating zero
electronic angular momentum and therefore an $\Omega$=0$^{+/-}$
character. As only a single rotational component was observed for
this line, no further assignment is possible. Nevertheless an
accurate analysis of the PA line shape can yield important
information about the atomic scattering state. More specifically,
the asymmetric broadening of the line towards red detunings
indicates the influence of the collision energy on the PA line shape
and can therefore be used as a measure for the relative temperature
of the two species. As discussed by Jones \textit{et
al.}~\cite{jones1999}, an accurate model of PA lines shapes can be
derived from Wigner's law. It states that for low collision energies
$\varepsilon$ the amplitude of the scattering wavefunction with
angular momentum $\ell$ scales as $\varepsilon^{(\ell+1/2)/2}$. With
the probability to have a collision with energy $\varepsilon$ given
by the Boltzmann factor $e^{-\varepsilon/T}$, the shape of a PA
resonance is then described by
\begin{equation}\label{eq:tempmodel}
W_l(f,f_0)=B\int_0^\infty
e^{-\varepsilon/T}\varepsilon^{(\ell+1/2)}L_\Gamma(f,f_0-\varepsilon)\,d\varepsilon,
\end{equation}
where $L_\Gamma(f,f_0)$ is the discrete energy Lorentzian with
natural linewidth $\gamma$ and central frequency $f_0$.
Fig.~\ref{fig:tempfit} shows the temperature broadened PA resonance
together with a fit of the model from Eq.~(\ref{eq:tempmodel})
taking only s-wave scattering ($\ell$=0) into account. The fit
yields realistic values for $T$=580(80)$\mu$K and $\Gamma$=10(3)MHz.
Adding higher partial waves did not improve the fit.

% Temperature broadened Omage=0 resonance and fit

%Modeling PA line shape: validity of model
Jones \textit{et al.} also derive a limit on the collision energy
$\varepsilon$, for which the Wigner law holds:
\begin{equation*}
\frac{h\varepsilon}{k_B} \ll \frac{E_{col}}{k_B}
\frac{2}{(0.6-A_0/R_0)^2}
\end{equation*}
where
\begin{equation*}
R_0=\left(\frac{2\mu C_6}{\hbar^2}\right)^{1/4} \textrm{ and
\hspace{2mm}} E_0=\frac{\hbar}{2\mu R_0^2}.
\end{equation*}
For $^7$Li$^{133}$Cs the reduced mass $\mu$ and the $C_6$ dispersion
coefficient~\cite{marinescu1994} are well known. However there are
only two experimental values for the scattering length $A_0$, namely
$A_{0,\textrm{singlet}}$=50(20)$a_0$ from high-resolution
spectroscopy of the singlet ground state
potential~\cite{staanum2007} and $A_{0,\textrm{eff}}$=180(40)$a_0$
from the thermalization of $^7$Li and $^{133}$Cs atoms in an optical
dipole trap~\cite{mudrich2002}. While the first value yields an
upper limit of roughly 0.8K well above our experimental conditions,
the latter one yields 1.7mK closer to the fitted temperature. As a
second condition the long-range wavefunction of the ground state has
to be linear near the outer turning point of the PA level. The PA
for the fitted resonance was performed at a detuning of more than
100cm$^{-1}$ from the asymptote where the excited potential energy
curve is still quite steep. Therefore the vibrational wavefunction
is very localized at the outer turning point and this second
condition should be well fulfilled. We conclude that if the relevant
scattering length for our experiment lies between
$A_{0,\textrm{singlet}}$ and $A_{0,\textrm{eff}}$, the derived
collision energy of $T$=580(80)$\mu$K should be a realistic estimate
for our experimental conditions.

%Centrifugal barriers in LiCs
This relatively high temperature has to be seen in relation to the
centrifugal barrier $E_\ell$ for collisions with higher angular
momentum $\ell$:
\begin{equation}
E_\ell=\left(\frac{\hbar^2 \ell (\ell + 1)}{3 \mu (2
C_6)^{1/3}}\right)^{3/2}.
\end{equation}
For $p$-wave scattering this yields $E_{\ell=1}$=1.6mK (barrier
peaked at 103$a_0$), for $d$-wave scattering $E_{\ell=2}$=8.5mK
(barrier peaked at 78$a_0$), and for $f$-wave scattering
$E_{\ell=3}$=24.7mK (barrier peaked at 66$a_0$). Our measured
collision temperature is well below the $p$-wave limit, so that one
can expect to observe dominantly $s$-wave scattering. Assuming a
Boltzmann distribution of kinetic energies, in roughly 6$\%$ of the
collisions the energy lies above the $p$-wave barrier, so that we
can expect a small contribution from $\ell$=1 but none from higher
partial waves.

% Molecular temperature
Using the above measured value of the collision temperature, we can
draw further conclusions about the temperature of the lithium atoms
and give an estimate on the temperature of the formed molecules. As
a simple approximation we assume an average collision angle between
lithium and cesium atoms of 90$^\circ$. The relative kinetic energy
is then given by
$E_{rel}=\frac{1}{2}\frac{m_{Li}m_{Cs}}{m_{Li}+m_{Cs}}(v_{Li}^2+v_{Cs}^2)$
and therefore the relative temperature
$T_{rel}=\frac{m_{Cs}T_{Li}+m_{Li}T_{Cs}}{m_{Li}+m_{Cs}}$ is
dominated by the lighter lithium. Solving this equation for T$_{Li}$
with T$_{Cs}$=250$\mu$K and T$_{rel}$=580$\mu$K yields
T$_{Li}$=600$\mu$K, which seems reasonable. For the center of mass
motion of the formed molecules
$E_{c.m.}=\frac{1}{2}(m_{Li}+m_{Cs})(\frac{m_{Li}\vec{v}_{Li}+m_{Cs}\vec{v}_{Cs}}{m_{Li}+m_{Cs}})^2$.
Therefore the molecular temperature
$T_{c.m.}=\frac{m_{Li}T_{Li}+m_{Cs}T_{Cs}}{m_{Li}+m_{Cs}}$ is
dominated by the heavier cesium. From the above determined
temperature T$_{Li}$ and T$_{Cs}$ we calculate a molecular
temperature of 270(60)$\mu$K.

\subsection{Photoassociation into the B$^1\Pi$ state}
\label{sec:paB1pi}

% The B1Pi State
The effort of searching for PA resonances especially at large
detunings can be substantially reduced when spectroscopic
information on the excited states of the system under investigation
is available. Most experimentally determined potential energy curves
are based on the observation of absorption and fluorescence between
bound levels, leading to a relative uncertainty on the order of
0.01\,cm$^{-1}$ in addition to an uncertainty of the dissociation
energy of the molecular ground state of up to 1\,cm$^{-1}$  and
therefore a corresponding uncertainty in the PA wavelength which
refers to the atomic asymptote. Once however a few levels have been
observed and assigned, this dissociation energy is known with high
precision and all other levels can be found easily. If no
spectroscopic data is available, \textit{ab initio} calculated
molecular potentials can be helpful. With up-to-date methods, the
typical uncertainty of the calculated potentials is on the order of
100\,cm$^{-1}$, while the vibrational spacings, related to the shape
of the potentials, are determined much more precisely. Therefore,
\textit{ab initio} potential curves provide a good estimate of the
typical number of expected PA resonances in a given energy range,
with an uncertainty for hitting one of them on the order of the
vibrational spacing. Once a vibrational progression is observed,
they are of importance for the final assignment and the prediction
of further vibrational levels.

% PA levels and lines
\begin{figure}[bt]
  \begin{center}
   \includegraphics[width=\columnwidth,clip]{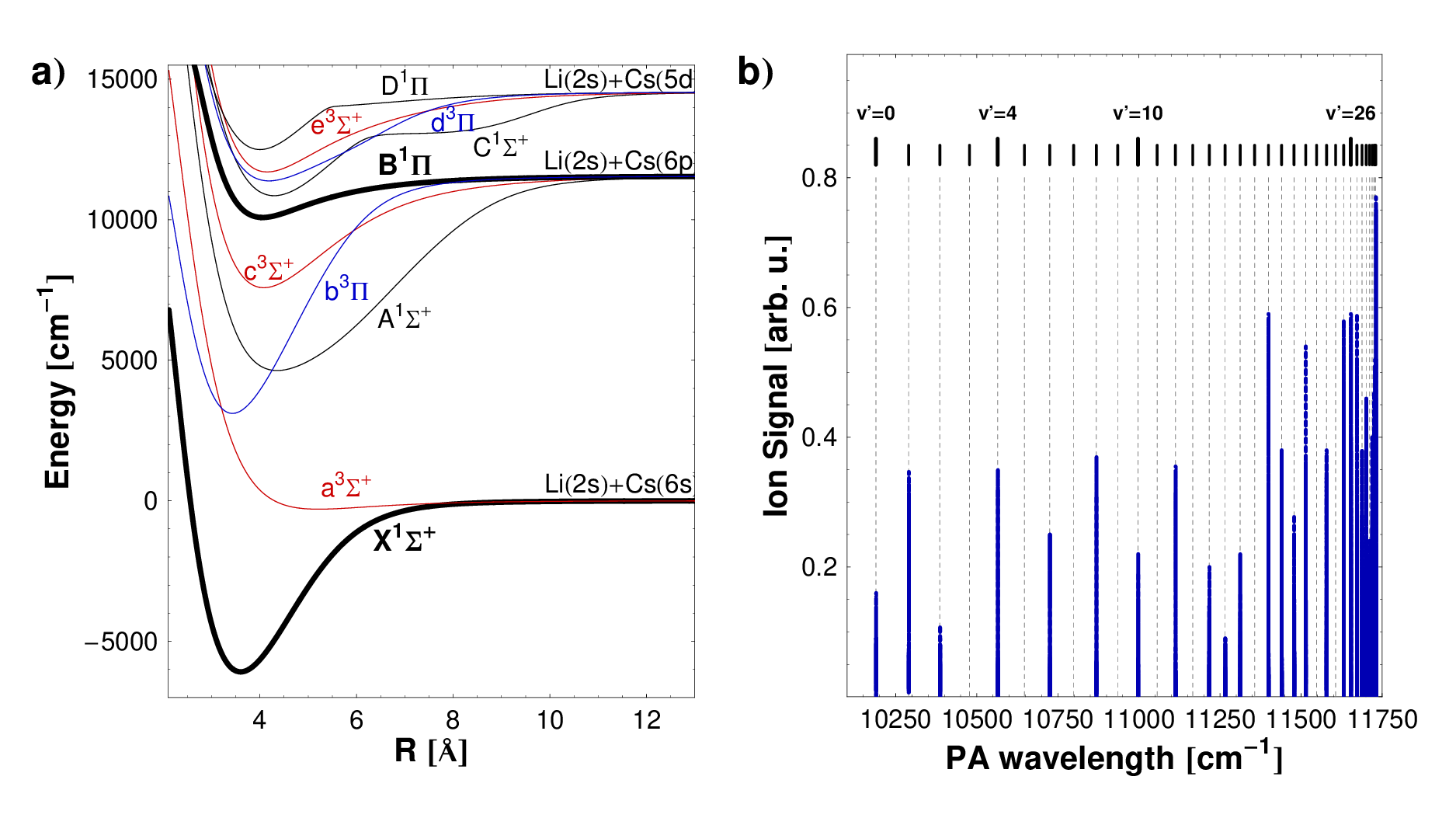}
   \caption{
   a) Potential energy curves for $\Sigma^+$ and $\Pi$ states of LiCs correlating to the three lowest asymptotes.
   The relevant states B$^1\Pi$ and X$^1\Sigma^+$ are marked bold.
   b) Overview over all observed vibrational levels in the B$^1\Pi$
   state.}
   \label{fig:paoverview}
  \end{center}
\end{figure}

In Fig.~\ref{fig:paoverview}\,a) \textit{ab initio} curves for
selected electronic states of LiCs correlating to the lowest three
asymptotes are shown (spin-orbit coupling is neglected in this
calculation). For two excited states, the B$^1\Pi$ and the D$^1\Pi$
state, Stein \textit{et al.} have published potential curves derived
from high resolution laser-induced fluorescence
spectroscopy~\cite{stein2008}. They report strong perturbations of
the rovibrational levels in the D$^1\Pi$ state due to spin-orbit
coupling to nearby triplet states. These perturbations reduce the
probability for decay into the lowest singlet state and also reduce
the accuracy of an extrapolation from the highly excited rotational
levels observed in Ref.~\cite{stein2008} to the lowest rotational
levels addressable by PA. In contrast to the D$^1\Pi$ state, the
B$^1\Pi$ state showed no significant perturbations, allowing for an
extrapolation of level energies to low rotational states with high
precision and making the spontaneous decay into the X$^1\Sigma^+$
ground state more likely. For the formation of ground state
molecules we therefore focused on the B$^1\Pi$ state. Extensive PA
scans were performed and PA resonances corresponding to
rovibrational levels in B$^1\Pi$ from the atomic asymptote down to
$v'$=0 at $\sim$1540cm$^{-1}$ detuning from this asymptote were
observed. For levels below $v'$=25 we found very good agreement with
the energies reported in Ref.~\cite{stein2008}. As the authors of
the cited work could only extrapolate from the last observed level
around $v'$=25 towards the asymptote, the predictions became less
accurate as we approached the asymptote. Here additional vibrational
levels up to $v'$=35 were found, yielding a data set for the
B$^1\Pi$ state with level energies ranging from the bottom of the
potential at $v'$=0 up to the last level below the asymptote, bound
only by $\sim$1\,GHz. A publication describing the complete B$^1\Pi$
potential energy curve is in preparation \cite{inprepA}. We note
that we did not observe line broadening due to predissociation in
the region between the fine structure asymptotes
Li(2$^2$S$_{1/2}$)+Cs(6$^2$P$_{3/2}$) and
Li(2$^2$S$_{1/2}$)+Cs(6$^2$P$_{1/2}$) as suggested by
Ref.~\cite{stein2008}. We are currently studying theoretically and
experimentally in how far predissociation could be relevant for
other electronic states correlated to the upper asymptote
Li(2$^2$S$_{1/2}$)+Cs(6$^2$P$_{3/2}$)~\cite{inprepB}.

In Fig.~\ref{fig:paoverview}\,b) an overview over all observed PA
resonances in the B$^1\Pi$ state is given. It is not straightforward
to compare the strength of the different PA lines, as not all PA
resonances could be detected at the same ionization wavelength and
the ion yield additionally strongly depends on the populated ground
state levels which vary between different PA resonances. However the
observed rate of molecule formation for PA into deeply bound
vibrational levels of B$^1\Pi$ is surprisingly high, as these
transitions occur at very short internuclear separations around
$R$=4{\AA} where the amplitude of the free scattering wavefunction
is already greatly reduced. It is noteworthy, that the Condon-points
of these deeply bound levels coincide with the inner turning point
of the a$^3\Sigma^+$ potential, which could be an important clue
towards a full explanation of the observed high PA rates.

\begin{figure}[tb]
  \begin{center}
   \includegraphics[width=0.7\columnwidth,clip]{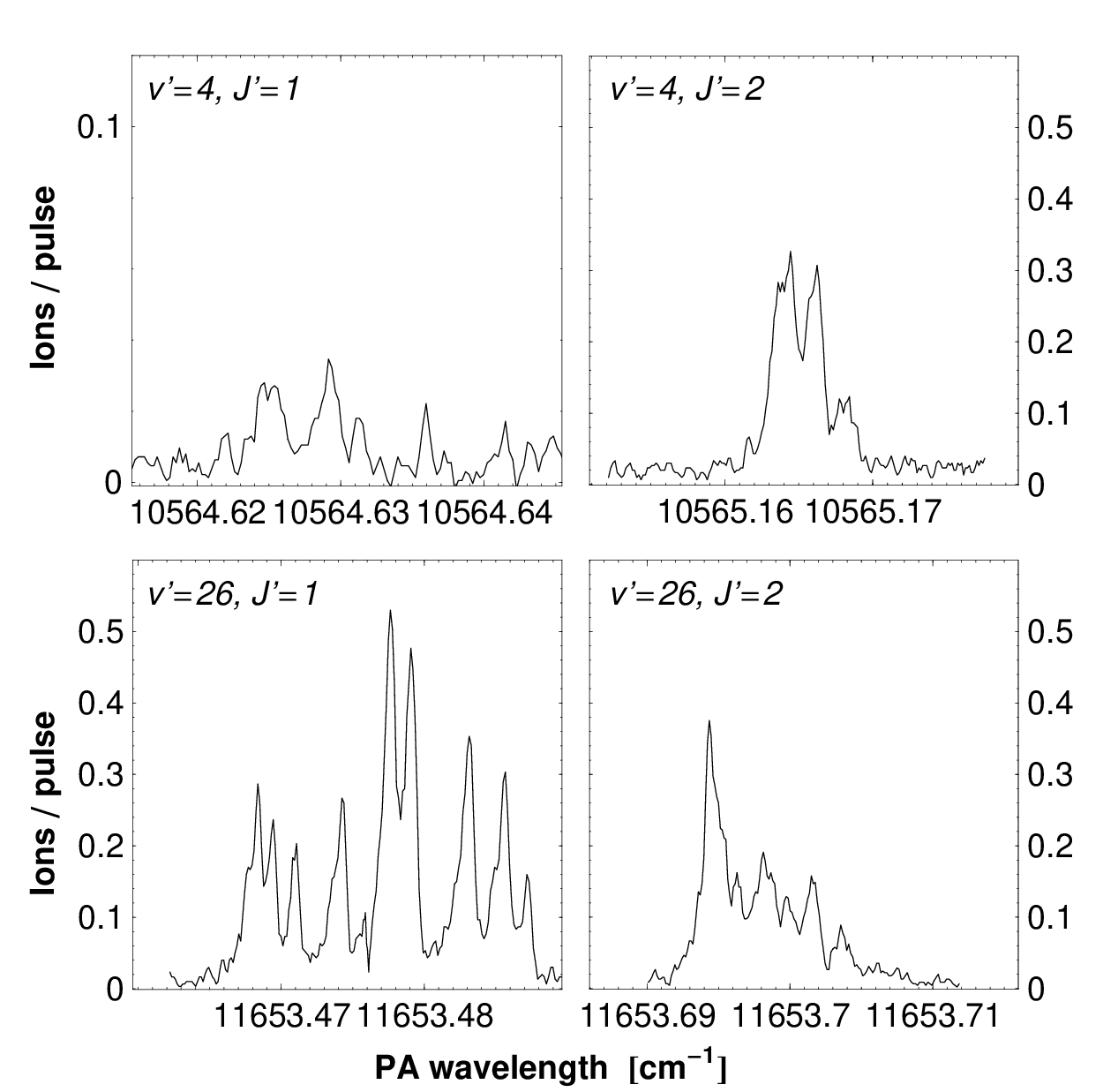}
   \caption{Photoassociation resonances for different rovibrational levels
   B$^1\Pi$,$v'$,$J'$. Abscissas and ordinates are on the same scale
   for all panels except for $v'$=4,$J'$=1, where the signal is
   scaled by a factor of 5 for visibility.}
   \label{fig:palines}
  \end{center}
\end{figure}

% PA line structures
The observed PA resonances in the B$^1\Pi$ state show a rich
substructure due to molecular hyperfine interactions as expected for
a state with electronic angular momentum. Exemplary resonances are
shown in Fig.~\ref{fig:palines}. The observed hyperfine splitting
generally decreases with increasing $J'$. We do not expect to
observe broadening of the lines due to stray fields, as the electric
extraction fields for the detection are switched on only 500$\mu$s
before the dye laser pulse, which ionizes all molecules that have
been produced within the last 20ms. Therefore the PA is performed
mainly under electric-field-free conditions. Also magnetic fields
should not influence the measured PA spectra. The magnetic field
gradient for trapping the atoms is applied constantly, but residual
fields in the trap center (where the gradient field reaches zero)
have been compensated to $<$1 Gauss.

% PA resonances for V'=4 and v'=26

%Rotational components
For low lying levels $v'$$<$25, rotational components $J'$=1 and 2
where observed, for levels $v'$$\geq$25 a weaker $J'$=3 component
and for the last bound levels rotational components up to $J'$=5
were visible. As the Condon-points for even the last bound levels
lie well within the centrifugal barrier for higher angular momenta
in the scattering continuum, the higher rotational components are
unlikely to be due to PA from atom pairs with higher angular
momentum. The observation of these higher $J$ components is probably
 due to the change of coupling between angular momenta as the
molecules become less tightly bound. From Fig.~\ref{fig:palines} one
can further learn, that while for the higher lying state the $J'$=1
component is the stronger one, for $v'$=4 the $J'$=2 component is
clearly dominant. This also indicates a change in the free-bound
coupling from PA at short distances around 5{\AA} to PA at larger
internuclear separations around 9{\AA} which is currently under
investigation.

\section{Calculated probabilities for decay into the X$^1\Sigma^+$ state }
\label{sec:pottheory}

%theory: calculation of FC factors
After PA, the formed excited molecules decay within a few
nanoseconds either into bound molecules or into free atom pairs. In
order to estimate the branching ratio between these two competing
channels and to predict the distribution of populated vibrational
ground state levels, we calculated Franck-Condon factors and
Einstein-A coefficients for transitions between the excited and the
ground state levels.

As mentioned before, accurate experimental potential energy curves
have been published for the X$^1\Sigma^+$ state\,\cite{staanum2007}
and for vibrational levels $v'\leq 25$ of the B$^1\Pi$ state
\,\cite{stein2008} of $^7$Li$^{133}$Cs. The authors of the latter
paper included our observed PA resonances in the fit describing the
full potential energy curve of the B$^1\Pi$ state~\cite{pashov2008},
yielding very accurate potential energy curves covering all binding
energies. We calculate rovibrational levels in these potentials
using the mapped Fourier grid Hamiltonian (FGH) method, which is
used to solve numerically the time-independent radial
Schr\"{o}dinger equation. A detailed description of the method may
be found elsewhere~\cite{kokoouline1999}. As expected, the resulting
term energies are in very good agreement with experimental
observations (residuals $\leq$0.02cm$^{-1}$). These transition
energies are used in the identification of ionization resonances
(Sect. \ref{sec:ionspec}) and depletion resonances (Sect.
\ref{sec:depletionpec}). The resulting wavefunctions are used to
calculate Franck-Condon (FC) factors for the bound-bound
transitions. It is worth mentioning, that for excited states
B$^1\Pi$,$v'$$\leq$26 the sum of FC factors with all vibrational
levels of the ground state is close to unity. Molecules in these
B$^1\Pi$ levels will therefore predominantly decay into bound
molecular ground state levels. We also note, that the authors of
Ref.~\cite{stein2008} report only for levels $v'\geq 17$ observation
of weak decay into the a$^3\Sigma^+$ lowest triplet state. One can
therefore expect, that levels B$^1\Pi$,$v'<17$ decay almost
completely towards X$^1\Sigma^+$ levels and that higher levels
exhibit only a weak additional decay channel towards a$^3\Sigma^+$
levels.

% Calculated level populations: dep. on dipole moment and v'
\begin{figure}[h]
  \begin{center}
   \includegraphics[width=0.7\columnwidth,clip]{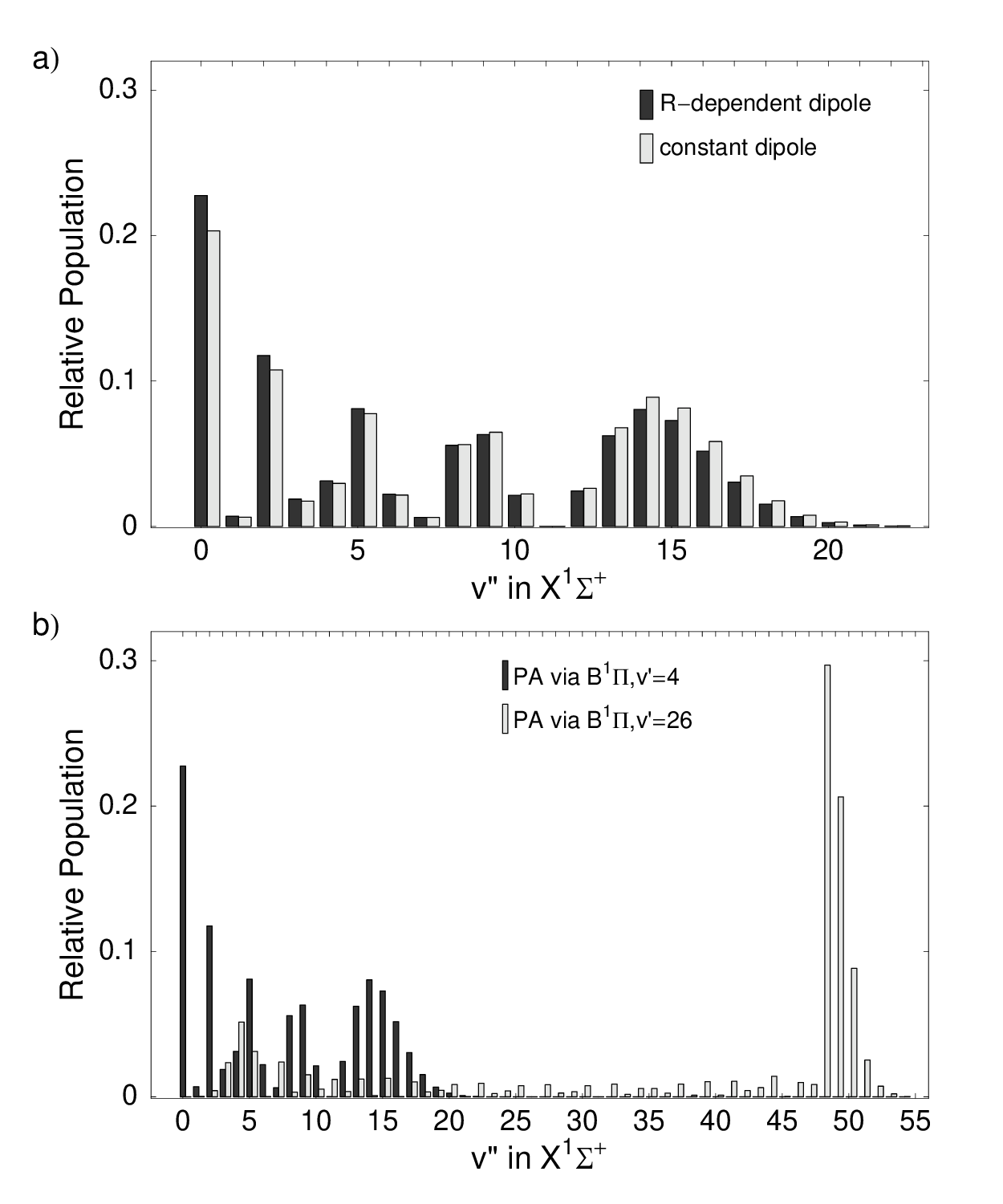}
   \caption{Calculated X$^1\Sigma^+$ level populations after PA, normalized to unity:
   a) influence of the R-dependence of the transition dipole moment
   on the expected distribution after PA via B$^1\Pi$,$v'$=4;
   b) comparison of the expected level population  after
   PA via B$^1\Pi$,$v'$=4 and $v'$=26.}
   \label{fig:vdist}
  \end{center}
\end{figure}

%calculation of ground state population
Using the approach described in Ref.~\cite{aymar2005}, the
$R$-dependent dipole transition moment $\mu(R)$ was calculated for
the electronic transition X$^1\Sigma^+$-B$^1\Pi$. The transition
dipole moment does not vary more than 10\% from a value of
9.9\,Debye over the relevant range of distances. By calculating the
Einstein $A$ coefficients for spontaneous decay
$A_{ij}=\frac{2}{3}\frac{\omega_{ij}^3}{\varepsilon_0 h
c^3}\left|\left\langle \Psi_i \big| \mu(R) \big|
\Psi_j\right\rangle\right|^2 $ with $\omega_{ij}$ being the
transition frequency and $\Psi_i$ the nuclear wavefunction of the
vibrational level $i$, one can derive the relative population of
 ground state vibrational levels after PA via different
excited states. In Fig.~\ref{fig:vdist}\,b) the normalized
population of X$^1\Sigma^+$ levels after PA via B$^1\Pi$,$v'$=4 and
B$^1\Pi$,$v'$=26 is compared. While in both cases low lying
vibrational levels are populated, we note that after PA via
B$^1\Pi$,$v'$=4, more than 20\% of the excited molecules should
decay into the lowest vibrational level $v''$=0.
Fig.~\ref{fig:vdist}\,a) shows the small influence of a calculation
with the R-dependent versus a constant dipole moment function for
this calculation.

\section{Ionization spectroscopy of X$^1\Sigma^+$}
\label{sec:ionspec}

The formation of ground state molecules is probed by ionizing those
molecules via resonantly-enhanced multiphoton ionization (REMPI) and
subsequent detection of the molecular ions (see
Sect.~\ref{sec:moleculesexp}). Once the formation of molecules by
the PA laser has been detected, a scan of the ionization laser
yields information about the distribution of populated vibrational
levels. Extracting this information requires accurate knowledge of
the involved ground and intermediate electronic states. In the case
of LiCs, deeply bound X$^1\Sigma^+$ levels $v''\leq 4$ can be
ionized via intermediate levels in the $B^1\Pi$ state which allows
us to assign observed ionization resonances to ground and
intermediate state vibrational levels.

\begin{figure}[p]
  \begin{center}
   \includegraphics[width=\columnwidth,clip]{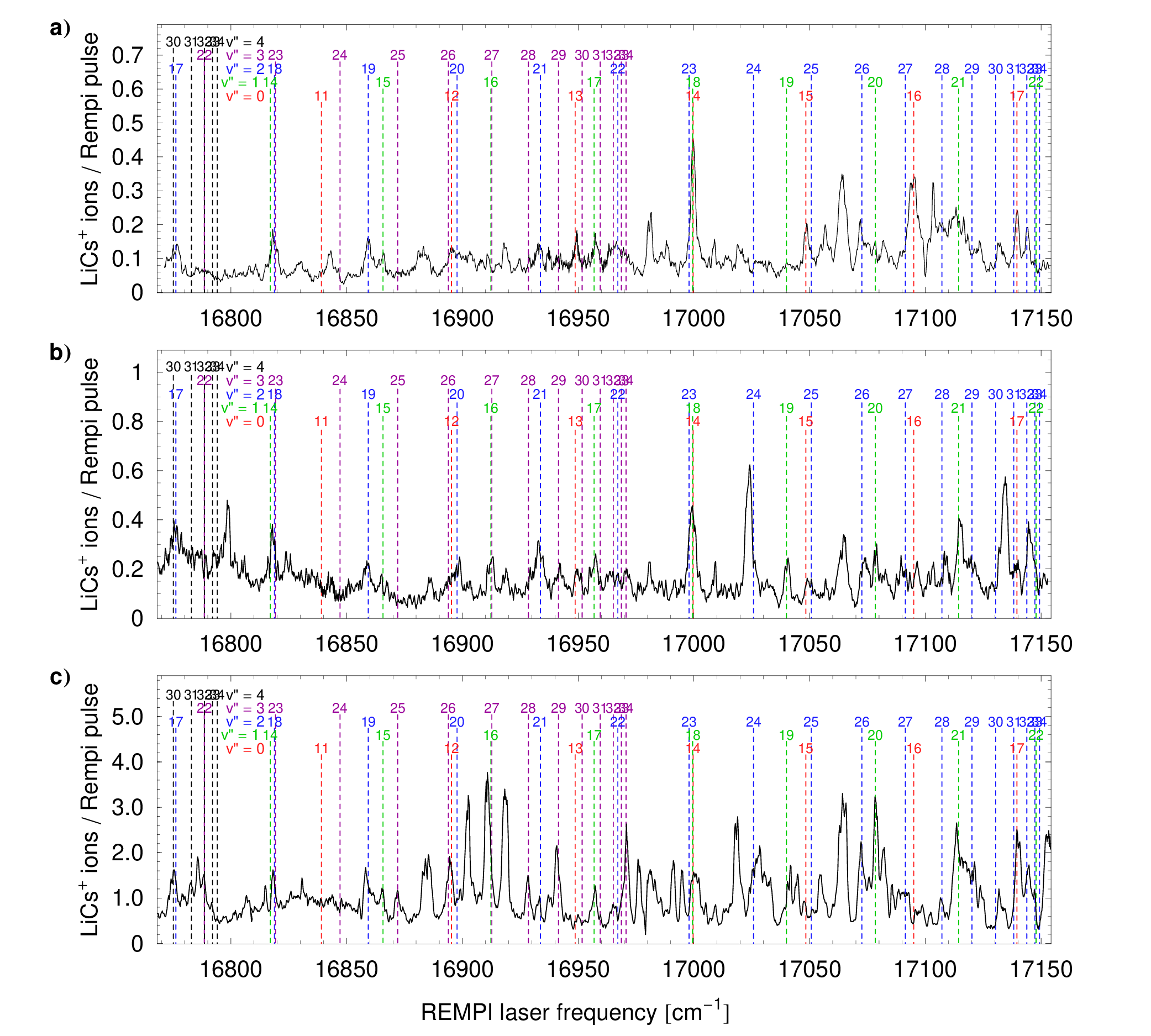}
   \caption{Resonantly-enhanced two-photon ionization spectra of ground state
    molecules after photoassociation via a) B$^1\Pi$,$v'$=4,$J'$=2, b) $v'$=10,$J'$=2,
    and c) $v'$=26,$J'$=2 over the same range of ionization energies.
    In the upper part of the graph, expected transitions series originating from all levels $v''$ in
    X$^1\Sigma^+$ to all levels $v'$ in B$^1\Pi$ are marked (note that
    this can not yield a full assignment of the observed lines,
    as different ionization probabilities and other intermediate states are neglected).
    Wavelengths calibrated with a High Finesse WS7 wavemeter with an
    accuracy $\ll 0.1$cm$^{-1}$. }
   \label{fig:rempigraph2}
  \end{center}
\end{figure}

%PA dependence of REMPI spectra
As discussed in the previous section, different PA resonances are
expected for different relative populations of ground state levels.
This can be seen very clearly in the three REMPI scans of
Fig.~\ref{fig:rempigraph2} covering the same range of ionization
energies after PA via different levels in the B$^1\Pi$ state.
Although the assignment of resonances to ground state levels is
often ambiguous and a number of unpredicted resonances indicate the
importance of intermediate states other than B$^1\Pi$, quite a few
clear progressions are visible. In Fig.~\ref{fig:rempigraph2}\,a),
resonances from X$^1\Sigma^+$,$v''$=0 to B$^1\Pi$,$v'$=13-17 and
X$^1\Sigma^+$,$v''$=2 to B$^1\Pi$,$v'$=17-21 can be identified; in
Fig.~\ref{fig:rempigraph2}\,b), in addition to the resonances
originating from $v''=2$, resonances from X$^1\Sigma^+$,$v''$=1 to
B$^1\Pi$,$v'$=14-21 are visible; and in
Fig.~\ref{fig:rempigraph2}\,c), in addition to molecules in $v''$=1
and $v''$=2, also molecules in $v''$=3 are ionized through
B$^1\Pi$,$v'$=22-29. The appearance of these ``ionization windows''
in the intermediate B$^1\Pi$ state demonstrates the importance of
the bound-bound transition dipole moment and the level-dependent
ionization probability of the B$^1\Pi$ state which lead to a
selection of ionization pathways.

% REMPI spectra from v'=4,v'=10 and v'=26

\begin{figure}[tb]
  \begin{center}
   \includegraphics[width=0.7\columnwidth,clip]{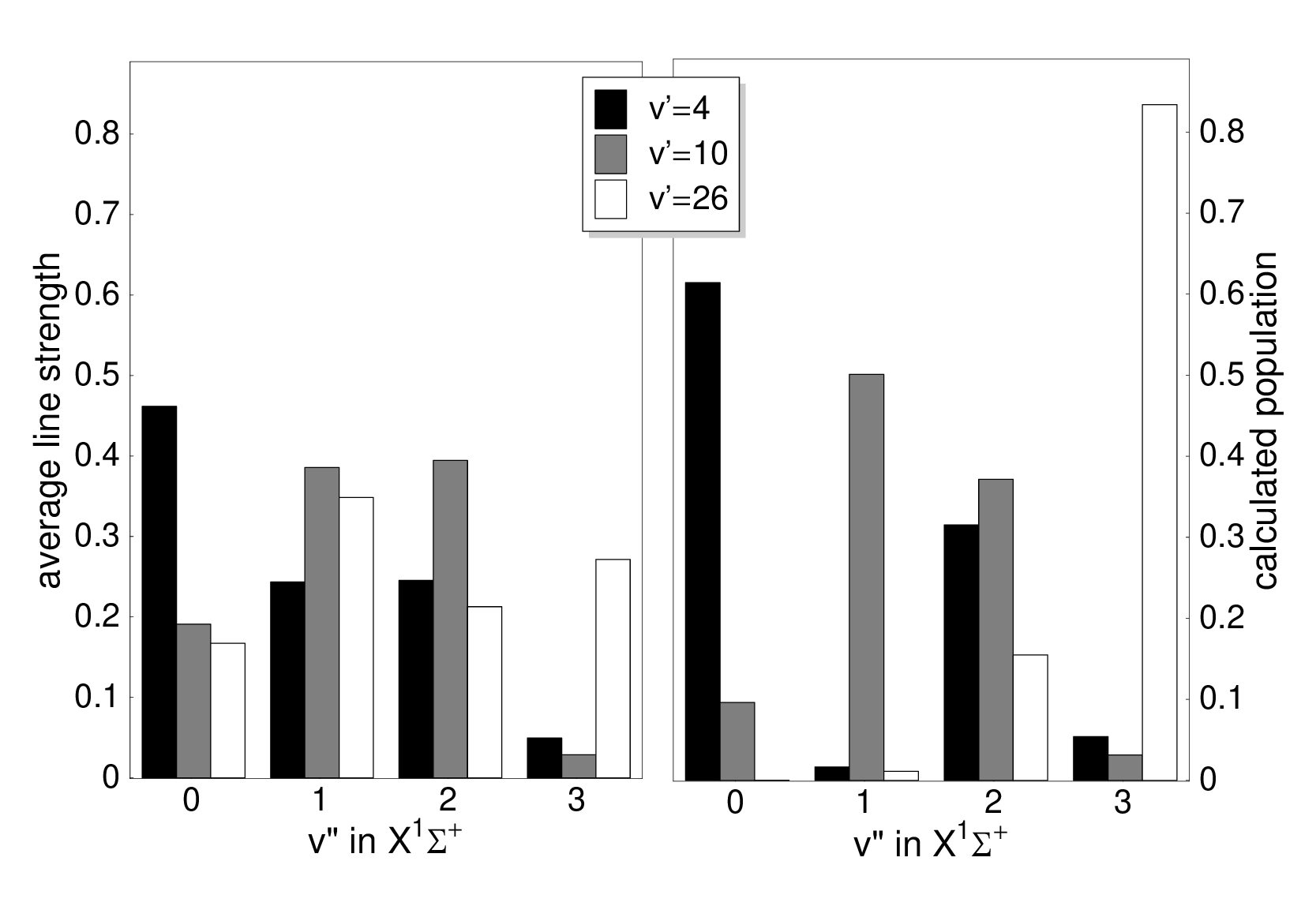}
   \caption{Comparison of detected ground state levels for different photoassociation resonances with predictions.
   Left panel: averaged peak value for the transition bands originating in X$^1\Sigma^+$,$v'$=0-3 as discussed in the text
   (normalized to unity for every PA resonance); Right panel:
    calculated ground state population, normalized to the summed population in levels $v'$=0-3 (see
    Sect.~\ref{sec:pottheory}).}
   \label{fig:measuredist}
  \end{center}
\end{figure}

% "measured" ground state population
The different and unknown ionization probabilities for different
intermediate B$^1\Pi$ levels as well as near degeneracy of many
transitions complicate an analysis of the relative populations of
vibrational ground state levels from the observed ionization
spectra. However, as shown in Fig.~\ref{fig:measuredist}, we can
established at least a qualitative agreement with the predictions of
Sect.~\ref{sec:pottheory}. In Fig.~\ref{fig:measuredist}, the
average peak value for each of the above identified transition bands
originating from ground state levels $v''$=0-3 is shown. While the
unknown ionization probabilities of intermediate levels impede the
direct extraction of the level population from the measured spectra,
we can still deduce the population strength of a single level after
PA via different excited levels and compare it with our prediction.
As can be seen in Fig.~\ref{fig:measuredist}, the relative ordering
of the line strengths for each level $v''$ is described correctly by
the theory, while the relative ratios show strong deviations. This
is very likely due to saturation and a non-optimal signal-to-noise
level.

% Comparing REMPI line strengths and expected ratios.

% depth of ion potential
The identification of the observed ionization resonances also yields
information about the lowest state of the ion LiCs$^+$, the
X$^2\Sigma^+$ state. The energy of two photons from
X$^1\Sigma^+$,$v''$=0 (via intermediate level B$^1\Pi$,$v'$=13)
leads to ionization at a detuning of -3385cm$^{-1}$ from the
Li(2\,$^2$S$_{1/2}$)+Cs$^+$($^1$S$_0$) threshold at
31406.71cm$^{-1}$\cite{moore1958}. Therefore the X$^2\Sigma^+$ state
of the ion has a depth of at least 3385cm$^{-1}$, favoring the
predictions of Von Szentpaly
($D_e$=3468cm$^{-1}$\,\cite{szentpaly1982}), Korek
($D_e$=3578cm$^{-1}$\,\cite{korek2006}), and Azizi
($D_e$=3564cm$^{-1}$\,\cite{Azizi2007}) over calculations of Patil
($D_e$=2662cm$^{-1}$\,\cite{patil2000}) and Bellomonte
($D_e$=2984cm$^{-1}$\,\cite{bellomonte1974}).

\section{High resolution spectroscopy of the vibrational ground state}
\label{sec:depletionpec}

%General description of depletion spectroscopy
After having identified vibrational ground state levels by selective
ionization of these levels, we can further probe the formed ground
state molecules by applying depletion spectroscopy~\cite{wang2007}.
While the Ti:Sa laser is stabilized to a PA resonance (see
Sec.~\ref{sec:paspec}) and the pulsed dye laser is kept on resonance
for the ionization of a certain ground state level, a narrow band
laser is used to optically pump the molecules out of this specific
level and therefore reduce the number of detected ions. As this
laser (Radiant Dyes cw dye laser, Rhodamin 6G pumped at 532\,nm) has
a bandwidth on the order of a few MHz, it allows one to resolve
rotational transitions between ground and excited molecular states.
The light from the laser is coupled into a fibre and, after being
collimated to a waist of 0.7mm, is aligned collinear with the
PA-laser by maximizing the depletion of the ion signal on resonance.
The frequency of this laser is measured with a wavemeter (Burleigh
WA-1000, relative stability 500MHz) and scans are monitored
additionally with a reference cavity (FSR=750MHz), which is actively
stabilized to an atomic cesium resonance via an offset-locked diode
laser. This high resolution spectroscopy of ground state molecules
allowed us to confirm assignments made in the interpretation of
REMPI spectra and to confirm the production of X$^1\Sigma^+$,$v''$=0
molecules after a single PA step~\cite{deiglmayr2008b}.

\begin{figure}[tb]
  \begin{center}
   \includegraphics[width=0.7\columnwidth,clip]{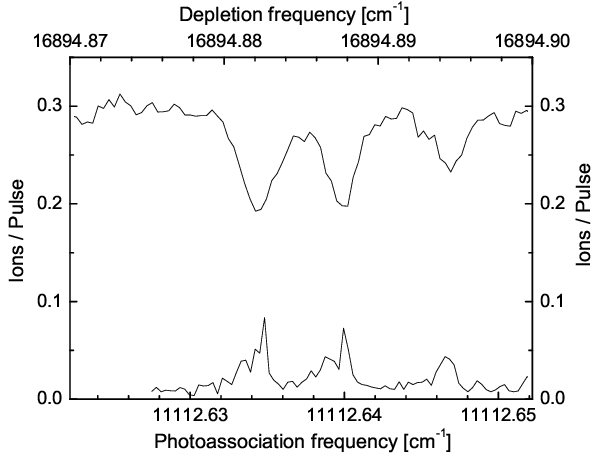}
   \caption{Depletion scan of X$^1\Sigma^+$,$v''$=0,$J''$=2 molecules via B$^1\Pi$,$v'$=12,$J'$=1 (upper trace)
   aligned to a photoassociation scan of the same excited level (lower trace). Depletion laser intensity 26\,mW/cm$^2$,
   photoassociation laser intensity 25\,W/cm$^2$. Molecules detected at different ionization wavelengths.}
   \label{fig:depletion}
  \end{center}
\end{figure}

% Determination of dissociation energy
The depletion spectroscopy makes it possible to gain further
information about the X$^1\Sigma^+$ potential energy curve. As shown
in Fig.~\ref{fig:depletion}, the B$^1\Pi$,$v'$=12,$J'$=1 level was
observed in PA spectroscopy starting from the
Li(2\,$^2$S$_{1/2}$,$F$=2)+Cs(6\,$^2$S$_{1/2}$,$F$=3) asymptote as
well as in depletion spectroscopy starting from
X$^1\Sigma^+$,$v''$=0,$J''$=2. Using the calculated rotational
constant $B_{v''=0}$=0.1874cm$^{-1}$ for the ground state, this
allows us to give a hyperfine averaged value for the binding energy
of the lowest vibrational level D$_0^X$=5783.53(3)\,cm$^{-1}$. This
agrees within the errorbars with the value of
D$_0^X$=5783.4(1)\,cm$^{-1}$ given by Staanum \textit{et
al.}~\cite{staanum2007}.

% Depletion spectroscopy

\section{An efficient route towards a dense gas of absolute ground state molecules}
\label{sec:outlook}

% summary
In this article we have discussed in detail the steps in the
formation and detection of ultracold polar LiCs molecules in the
vibrational ground state. We have shown that a single
absorption-emission cycle is sufficient to produce the molecules in
the lowest vibrational state. Previously we have demonstrated also
the formation of ultracold LiCs molecules in the rotational and
vibrational ground state
X$^1\Sigma^+$,$v''$=0,$J''$=0~\cite{deiglmayr2008b} in a single step
of PA. In the future we will repeat these experiments with lithium
and cesium atoms trapped in a quasi-electrostatic trap (QUEST)
formed by a focused CO$_2$ laser at 10.6$\mu$m. This very large
detuning from any atomic or molecular resonance makes the absorption
of a trap laser photon by LiCs molecules very unlikely: a transition
from deeply bound molecules into the first excited state requires
roughly ten photons from the trap laser. Due to the permanent dipole
moment of the polar molecules, transitions within an electronic
state are in principle also possible; however, they are very
unlikely due to extremely small Franck-Condon overlaps for such
transitions.

From a calculation of the static polarizability of LiCs molecules in
X$^1\Sigma^+$,$v''$=0~\cite{deiglmayr2008a} and previous results
obtained with this setup~\cite{mosk2001}, we expect a trap depth on
the order of 600$\mu$K for the ground state molecules in $v''$=0.
The calculation shows, that the trap depth increases for higher
vibrational levels, so that all produced ground state molecules are
expected to be trapped. While the molecules in $v''$=0,$J''$=0 are
expected to be stable against collisions with other atoms and
molecules in the same state, collisions with vibrationally excited
molecules can lead to a loss of $v''$=0 molecules. It will therefore
be useful to apply the PA laser only as long as the number of atoms
in the trap significantly exceeds the number of produced molecules,
in which case any molecule would most likely collide with a trapped
atom and not with a molecule. As demonstrated for Cs$_2$+Cs
collisions~\cite{staanum2006,zahzam2006} and RbCs+Cs/Rb
collisions~\cite{hudson2008}, such collisions lead to a fast removal
of vibrationally excited molecules from the trap. After an
accumulation time on the order of seconds, one would end up with a
relatively pure sample of molecules in the rovibrational ground
state. Optimistically, 20\% of the trapped atoms could be
transferred into molecules in the rovibrational ground state.
Starting from atomic densities on the order of 10$^{11}$cm$^{-3}$
and atomic temperatures around 20$\mu$K~\cite{mudrich2002}, one can
therefore hope to reach molecular samples with densities on the
order of 10$^{10}$cm$^{-3}$ and temperatures also around 20$\mu$K.
While this is far away from a degenerate gas of polar molecules, the
long-range dipolar interactions will already at these densities lead
to a strongly interacting system. After stabilizing the polar gas
against inelastic collisions due to these interactions,
\textit{e.g.} via a ``blue shield'' for molecules
~\cite{gorshkov2008} or by preparing aligned, 2-dimensional samples
in combined static electric fields and strong laser
fields~\cite{deiglmayr2008a}, the realization of a new quantum phase
like a dipolar crystal~\cite{pupillo2008} seems feasible.\\

%Acknowledgements
\textit{Acknowledgements}: We thank S.D. Kraft and P. Staanum for
contributions at the early stage of the experiment. We also thank E.
Tiemann and A. Pashov for providing experimental LiCs potentials
before publication and, together with J. Hutson, for fruitful
discussions. This work is supported by the DFG under WE2661/6-1 in
the framework of the Collaborative Research Project QuDipMol within
the ESF EUROCORES EuroQUAM program. JD acknowledges partial support
of the French-German University. AG is a postdoctoral fellow of the
Alexander von Humboldt-Foundation.

%The references should start on their own page.
\clearpage

\bibliography{mixtures}

\end{document}